# Correlation between structure and dynamics of $CO_2$ confined in Mg-MOF-74 and the role of inter-crystalline space: A molecular dynamics simulation study

I. Dhiman[a,*], M. C. Berg[b], David R. Cole[c], Siddharth Gautam[c]

[a]Centre for Energy Research, Budapest, 1121 Hungary

[b]Jülich Centre for Neutron Science at Heinz Maier-Leibnitz Zentrum, Forschungszentrum Jülich GmbH, Garching, Germany

[c]School of Earth Sciences, The Ohio State University, 275 Mendenhall Laboratory, 125 S Oval Drive,

ü*Email: *indu.dhiman@ek-cer.hu*

**Abstract**
________________________________________________________________________________
Mg-MOF-74 is a metal-organic framework (MOF) that exhibits a high capacity for $CO_2$ adsorption. Given the importance of $CO_2$ confinement in Mg-MOF-74 for capture and storage applications, it is important to understand the structural and dynamical behavior of $CO_2$ in Mg-MOF-74 pores. While most molecular simulation studies use ideal single crystal models of nano-porous substrates, the existence of inter-crystalline space has been shown to have profound effects on the sorption, structure and dynamics of the adsorbed fluid. To address these issues, we report a molecular dynamics simulation study at 300 K, of $CO_2$ confined in several models of Mg-MOF-74 with systematically inserted inter-crystalline spacing of different widths. Both structural and dynamical behavior of $CO_2$ is studied in 5 models of Mg-MOF-74, each at 4 different loadings. Six strong sites of $CO_2$ adsorption are found at the periphery of the pores of Mg-MOF-74 in addition to a relatively weak adsorption at the center of the pore. On insertion of inter-crystalline spacing, additional six sites of strong adsorption are seen in the inter-crystalline space close to the pore opening. These additional sites delocalize as the inter-crystalline space is widened and the population of guest molecules adsorbed at the pore center grows at the expense of peripheral population. This redistribution of guest molecules has important implications for their dynamics. While in the model without inter-crystalline space, translation motion is found to be slower at higher loadings, as wider inter-crystalline space is introduced, anomalous loading dependence of translational diffusivity is observed – a direct consequence of the redistribution of guest populations. In general, inserting inter-crystalline spacing is found to enhance both translational as well as rotational motion of the guest molecules. The inter-relation of structure and dynamics of $CO_2$ molecules confined in Mg-MOF-74 and their variation with inter-crystalline space as reported here provide valuable insight to carbon capture and storage.
________________________________________________________________________________

## 1. Introduction

One of the most critical environmental challenges today is the rise in atmospheric $CO_2$ levels, one of the main components contributing towards climate change.[1, 2] Currently, the main source of this rise is combustion of fossil fuels, such as coal, natural gas and petroleum etc.[1-7] With growing demand for energy, it is clear that containment of $CO_2$ levels in environmental-friendly ways is one of the greatest challenges today. One of the initial methods utilized to reduce $CO_2$ emission, particularly from flue gas, has been amine scrubbing. However, amine scrubbing is not an energy efficient technique, and leads to harmful ramifications.[3, 4] In recent years, technological advancement has led to more efficient and environmental/ecological friendly ways of $CO_2$ separation via solid adsorbents. Towards this, potential candidates investigated so far as solid absorbents are mostly silicates,[8-13] zeolites,[14-20] carbon-based materials,[21-25] or metal–organic frameworks (MOF).[26-39] In particular, MOFs are some of the most promising candidates for these applications and have gained significant attention recently for their high $CO_2$ adsorption capacity. Various possible applications of MOFs extend to areas such as gas - liquid separation, gas separation, gas storage, catalysis, drug delivery, biotechnology etc.[40-49]

Crystalline MOFs comprise metallic centers connected by organic linkers.[50] Compared to other absorbent materials MOFs have extremely large internal surface area. This makes them one of the most promising, and highly diverse customizable materials. This implies that by tuning the combination of the organic parts and metal ions, properties of MOFs relevant to applications can be controlled. Among all the MOFs investigated so far, Mg-MOF-74 has been one of the most efficient in capturing $CO_2$ due to its high affinity and capacity. Previous studies indicate this behavior is due to the presence of coordinated unsaturated metal sites (CUS) or open metal sites (OMS).[51-57] A density functional theory (DFT) study reported by Park et al. highlights this aspect of Mg-MOF-74.[58] This study demonstrated that this high affinity towards $CO_2$ is a result of the strong interaction between the $CO_2$ guest molecules and OMS. This is linked to the strong electrical field arising from the electronic density of the $Mg^{2+}$ ions, resulting in the formation of a coordination bond between the empty orbitals of Mg ions and $CO_2$ electrons.

Nano-porous adsorbent materials (such as MOFs) usually occur in the form of powder with grains and particles separated from each other by inter-particle spacings. These grains in turn are made up of several crystallites that may be separated from each other with an inter-crystalline spacing.[59-63] While the simulations are mostly carried out on idealized models of the porous adsorbents, the existence of inter-crystalline spacing is mostly ignored. A recent study by Gautam and Cole showed the importance of this inter-crystalline spacing in the absorption of $CO_2$ in silicalite and Mg-MOF-74 using Monte Carlo simulations.[62, 63] It was observed that the presence of inter-crystalline spacing enhances the $CO_2$ adsorption. The effects of inter-crystalline spacing on the diffusivity of guest molecules adsorbed in zeolites BEA and Na-Y have also been reported.[59, 60] It was observed that incorporating inter-crystalline spacing in the substrate models lead to a better agreement between the diffusivities obtained from the simulations and experiments. While inter-crystalline space was included in the substrate models in these works, no systematic variation of the inter-crystalline space width was carried out. In the study reported by Gautam and Cole the effects of inter-crystalline spacing on adsorption were studied by systematically varying the width of inter-crystalline space.[62, 63] Given the importance of the system of $CO_2$ confined in Mg-MOF-74 for $CO_2$ storage studies and the need to systematically study the effects of inter-crystalline spacing on the behavior of adsorbed species, we carried out an MD simulation study on the structure and dynamical behavior of $CO_2$ confined in Mg-MOF-74 with and without inter-crystalline spacing of different widths. In particular, this study addresses the questions: how does a difference in the adsorbed amounts of $CO_2$ affect its dynamics within Mg-MOF-74? and what role can inter-crystalline spacing play in the diffusivity of $CO_2$ within the pores and in the inter-crystalline spacing between Mg-MOF-74 crystallites? This study provides important information on the dynamics of $CO_2$ in realistic samples (powder like vis-a-vis perfect crystalline) of Mg-MOF-74, which is essential for understanding the $CO_2$ storage capacity of this adsorbent.

## 2. Simulation Details

Classical molecular dynamics simulations are carried out using DL_POLY_4 v5.0.0.[64] The optimization of Mg-MOF-74 unit cell parameters have been reported by Yazaydin et al.[65] The CIF file for the same is taken from the software package RASPA repository.[66] A unit cell is replicated to 2 × 2 × 6 using VESTA visualization software. For consistency with previous study by Gautam et al. the nomenclature used in that study is retained in the current work.[63] S0 denotes the simulation cell without any inter-crystalline spacing, therefore constituting a crystallite. The simulation cell S0 is shown in Figure 1(a). Thereafter, each simulation cell is modified by adding an empty space of 1.5, 2.5, 5.0, or 10 Å to S0 and the simulation cells thus obtained are denoted as S1, S2, S3, and S4 (Figure 1(b) and (c)), respectively. The respective empty space is added along all the six sides of S0. Periodic boundary conditions are applied along all the directions. As a result, these simulation cells constitute infinite number of crystallites with respective separation of 3, 5, 10, or 20 Å within inter-crystalline spaces.

For the S0 simulation cell, $CO_2$ molecules are added at loading values of n = 5, 10, 15 and 20 molecules per unit cell (mpuc) leading to total number of $CO_2$ molecules of 120, 240, 360 and 480 respectively. These loadings for S0 are equivalent to a gas partial pressure of 0.3 - 3 bar as inferred from the adsorption isotherms obtained earlier for S0.[63] Both $CO_2$ as well as Mg-MOF-74 were simulated as rigid bodies.

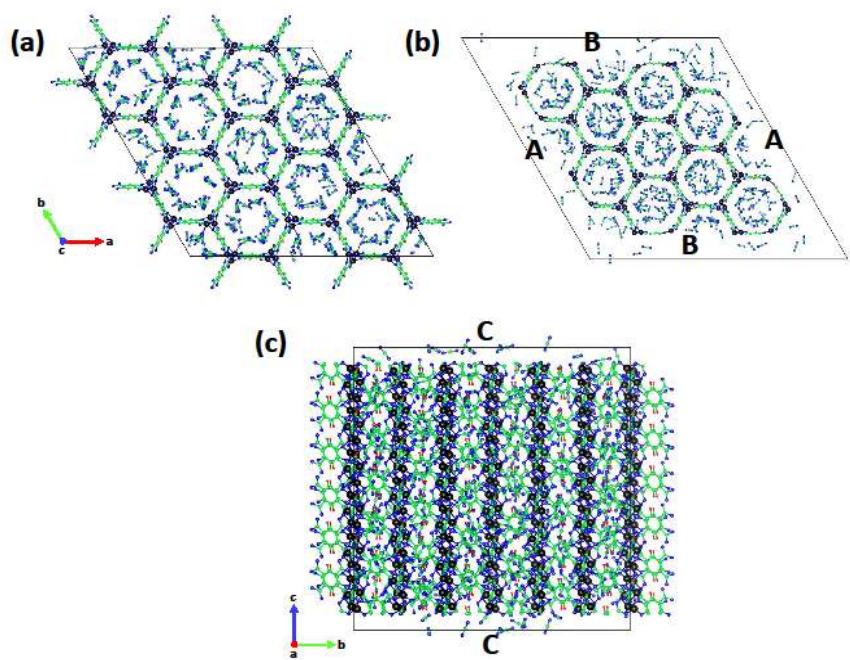

**Figure 1:** (a) Simulation unit cell S0 in the X-Y plane, without inter-crystalline spacing. The unit cell boundary is marked with the black outline shown in the figure. (b) Simulation unit cell S4 in the X-Y plane, with inter crystalline spacing of 10 Å introduced along all the six surfaces. Also shown are the crystallite surfaces A and B. The Inter-crystalline spacing of 5 Å is added to these surfaces individually to obtain unit cell A2 and B2, respectively. (c) Simulation unit cell S4 in the Y-Z plane. Also marked is the crystallite surface C, used to build a simulation cell C2.

We further expanded the simulations to investigate the effect of surfaces along different planes exposed. Specifically, crystallite surfaces parallel to the cell vectors a, b, and c are identified as A, B, and C, respectively. The corresponding surfaces are shown in Figure 1(b) and (c). To compare the effect of individual surfaces exposed, inter-crystalline spacing of 5 Å along different surfaces is introduced in simulation cell S0. The new formed simulation cells are termed as A2, B2 and C2 in accordance with the respective surfaces exposed. The respective simulation cells are prepared with empty spaces introduced only on two opposite surfaces, whereas the remaining four surfaces are devoid of any empty space.

The simulations are performed using NVT ensemble for 6 ns with trajectories saved at every 0.01 ps step. The equilibration of the system is assured when the fluctuations in energies and temperatures are below 5%. This happened after the first 1 ns of simulation. The resulting trajectories are unwrapped using a visual molecular dynamics (VMD) software tool.[67] Calculations on long range interactions are carried out using 3D Ewald sum. Periodic boundary conditions have been implemented along all the directions. TraPPE-UA convention is used with cut off of 14 Å towards all the interactions. An integration step of 1 fs is used for all the simulations.

## 3. Results

### 3.1 Structural Behavior

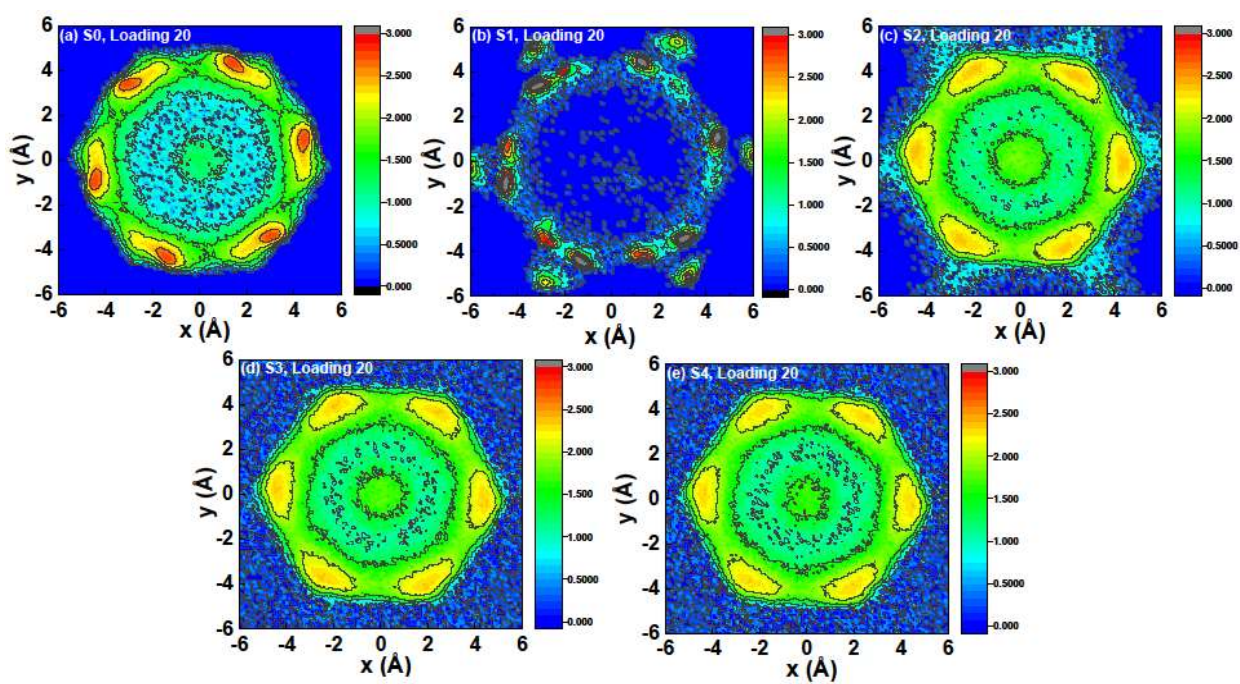

**Figure 2:** Distribution of carbon atoms belonging to $CO_2$ molecules for loading n = 20 mpuc in Mg-MOF-74 in (a) S0, (b) S1, (c) S2, (d) S3, and (e) S4. The Z-intensity represents log ($N$ + 1), where $N$ is the number of $CO_2$ molecules (carbon atoms) residing at a given location (projection in the X–Y plane) for the entire production run time (10 001 frames).

To visualize the localization of $CO_2$ molecules, heat maps showing the distribution of carbon atoms belonging to $CO_2$ molecules in the XY plane for S0 – S4 corresponding to $CO_2$ loading of 20 mpuc are shown in Figure 2. The Z-intensity in this figure is log ($N$ + 1), where $N$ is the number of carbon atoms belonging to $CO_2$ molecules residing at a given location (projection in the X–Y plane) for the entire production run time. The intensity in this distribution plot is related to the probability of finding the carbon atom belonging to a $CO_2$ molecule at a given location and time. Distinct difference in $CO_2$ molecular distribution for S0, S1 and S2-S4 inter-crystalline spacings are observed. In S0, six high intensity locations can be seen on the periphery of the pore. These are the preferred sites of adsorption where a $CO_2$ molecule is more likely to be localized. In addition, the pore center has a moderately high intensity region too, suggestive of a layering structure at higher loadings where a central layer of adsorption begins to form once the peripheral sites are saturated. At lower inter-crystalline spacing S1, while the six adsorption sites are retained, the intensity in the central region of the pore as seen in S0 disappears. In addition, six high intensity regions can be seen that appear to be outside the pore but are close to the six strong adsorption sites. These additional sites of adsorption exist in the inter-crystalline space close to the pore opening. As the inter-crystalline space gets wider in S2-S4, these additional six sites of adsorption start to delocalize, and the central region of moderately high intensity reappears. Within a pore, regularly spaced localized sites of adsorption are observed in the axial direction, in XZ and YZ planes (not shown).

3.2 Translation Dynamics

To understand the translation dynamics of $CO_2$ molecules in Mg-MOF-74, mean square displacement (MSD) as a function of time is calculated. The overall MSD variation with time for the loadings of 5, 10, 15, and 20 mpuc for S0, S1, S2, S3 and S4 gaps is shown in Figure 3. For gap S0 in Figure 3(a), linear dependence of MSD is observed as a function of time for all the loadings. With increasing loading, a reduction in MSD values is observed. For systems S1 and S2, in Figure 3(b) and (c) respectively, with higher inter-crystalline spacing similar loading behavior in MSD is observed. However, this loading dependence is reversed when a wider inter-crystalline spacing is added in S3 and S4.

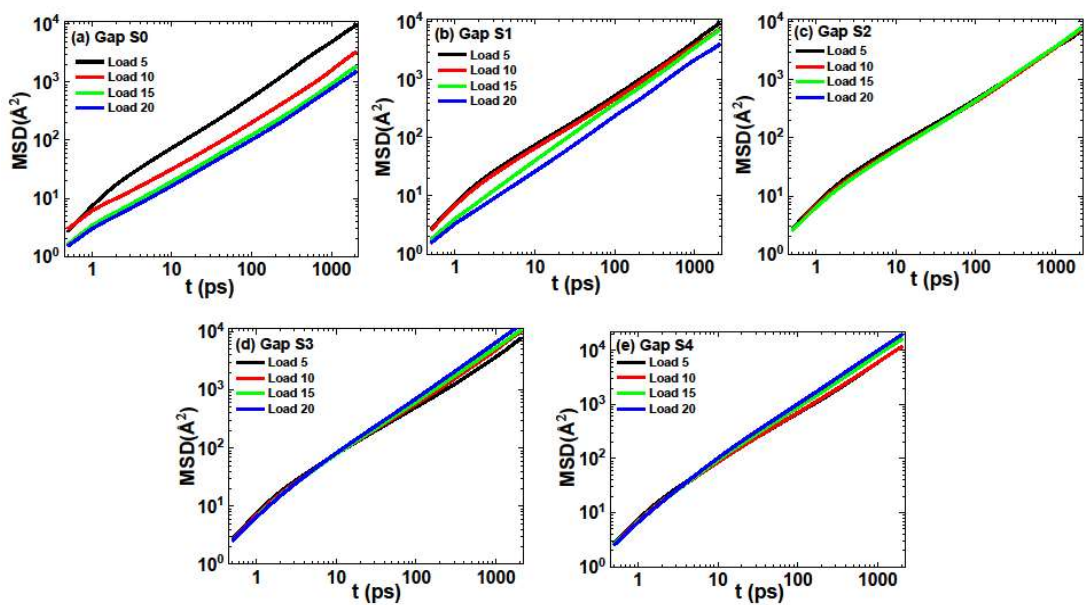

**Figure 3:** Time dependent mean square displacement (MSD) as a function of loadings, n = 5, 10, 15, and 20 mpuc for (a) S0, (b) S1, (c) S2, (d) S3, and (e) S4.

We also calculated the MSD components in a direction parallel and perpendicular to molecular axis and with respect to the Cartesian X-, Y- and Z- directions. No significant change in MSD behavior in either a direction parallel or perpendicular direction to the molecular axis is observed as a function of loading for all the gaps. MSD as a function of time for all the inter-crystalline spacings for loading n = 20 mpuc with respect to the X-, Y- and Z- directions is shown in Figure 4. The figures shown here for the respective inter-crystalline spacings are also representative for the lower loadings, n = 5, 10, and 15. For S0 the MSD along Z- direction, as compared to X- and Y- direction, exhibits higher values. With absence of an inter-crystalline gap as in the case of S0, only the Z- direction facilities the movement of $CO_2$ molecules.

As inter-crystalline gaps are introduced in MOF, molecules also become relatively free to move along X- and Y- direction in addition to Z- direction. This leads to a relative increase in MSD values along both X- and Y- directions. Also, as expected with increasing the gaps from S1 to S4, the difference in MSD between X-, Y-, and Z- direction is reduced.

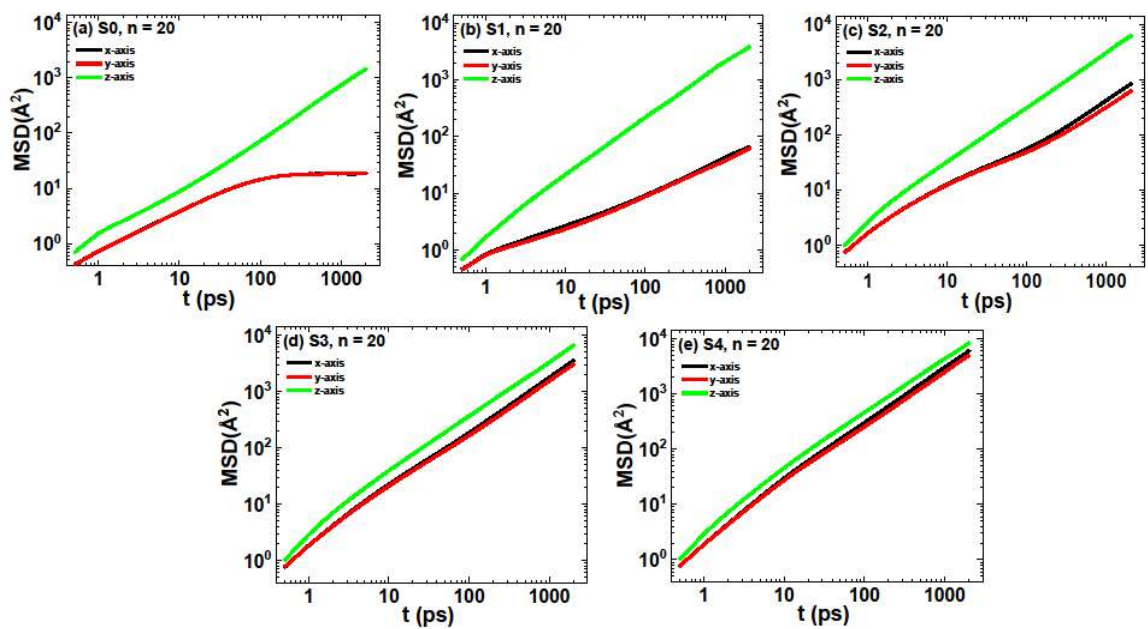

**Figure 4:** Mean square displacement (MSD) for loading, n = 20 mpuc for (a) S0, (b) S1, (c) S2, (d) S3, and (e) S4. The plots shown here are representative for the lower loadings of n = 5, 10, and 15 mpuc, and respective inter-crystalline spacings.

Diffusion coefficient as a function of loading for all the gaps are calculated using the Einstein relation for the time scales above 100 ps (approx.), well beyond the ballistic region that lasts for only the first few picoseconds. The linear time dependence of MSD in this region is ascribed to the diffusive motion of molecules. The diffusion coefficient for the overall MSD (Figure 5(a)) for gap S0 is reduced as a function of loading. This behavior is ascribed to the crowding of molecules with increased loading, restricting the motion of $CO_2$ molecules. Similar behavior is observed for S1 and S2. A transition is observed at S3 and S4, wherein the diffusion coefficients as a function of loading exhibit an increase. This occurs due to additional space available for $CO_2$ molecules in the center of gaps as we shall describe in detail in the discussion section.

The diffusion coefficient as a function of loading along X-, Y- and Z- direction are shown in Figure 5(a), (b), and (c), respectively. Like the overall D, the diffusion coefficient along X- and Y- directions increase with increasing loading and with inter-crystalline spacings.

Interestingly for mobility along Z-axis we observe D values reduce with increasing loading for gaps S0 and S1. Further increase in gaps from S2 to S4 yields D values exhibiting an increase with loading. This trend is similar to one described for D values obtained for overall MSD in Figure 5(a). This implies that the overall behavior observed is dominated by the motion of $CO_2$ molecules along Z- direction, correlated to the MOF structure with less confinement along the Z-direction.

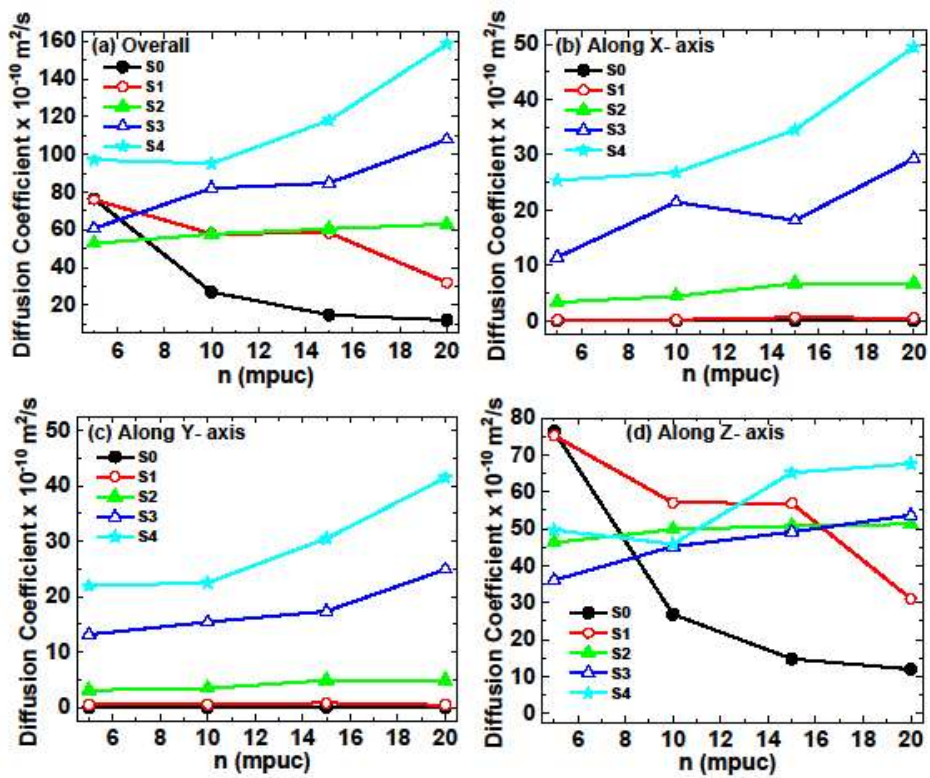

**Figure 5:** Direction specific diffusion coefficient as a function of loading for (a) overall (b) X- (c) S2, (d) S3, along X-, Y- and Z-direction as a function of loading.

3.3 Rotational Dynamics

Rotational behavior of molecules is studied by calculating the rotational correlation function (RCF), $C_R(t)$ using the equation shown below,

$$C_R(t) = \langle \boldsymbol{u}(t + t_0).\boldsymbol{u}(t_0)\rangle$$

A unit vector ($\boldsymbol{u}$) here is assigned to the molecular axis. The angular brackets in the equation represents an average over all the molecules, and time of origin $t_0$. The RCF is related with dipole correlation functions and can be experimentally obtained using dielectric or infra-red spectroscopy. In Figure 6(a), the time dependence of RCF for S0, at different loadings is shown. At lower loadings, a faster decay of RCF is observed, implying faster rotational motion of molecules. Over a longer timescale of a few tens of picoseconds the RCF is completely decayed. In contrast, at higher loadings RCF decay is relatively slow, owing to the crowding of the molecules restricting the rotation. Similar behavior is observed for S1, depicted in Figure 6(b). As the inter-crystalline spacing is further increased to S2, S3 and S4, shown in Figure 6(c), (d), and (e) respectively, the RCF displays a similar decay rate for all the loadings. This may imply that for higher spacings molecules are no longer constrained, hence favoring free rotational motion.

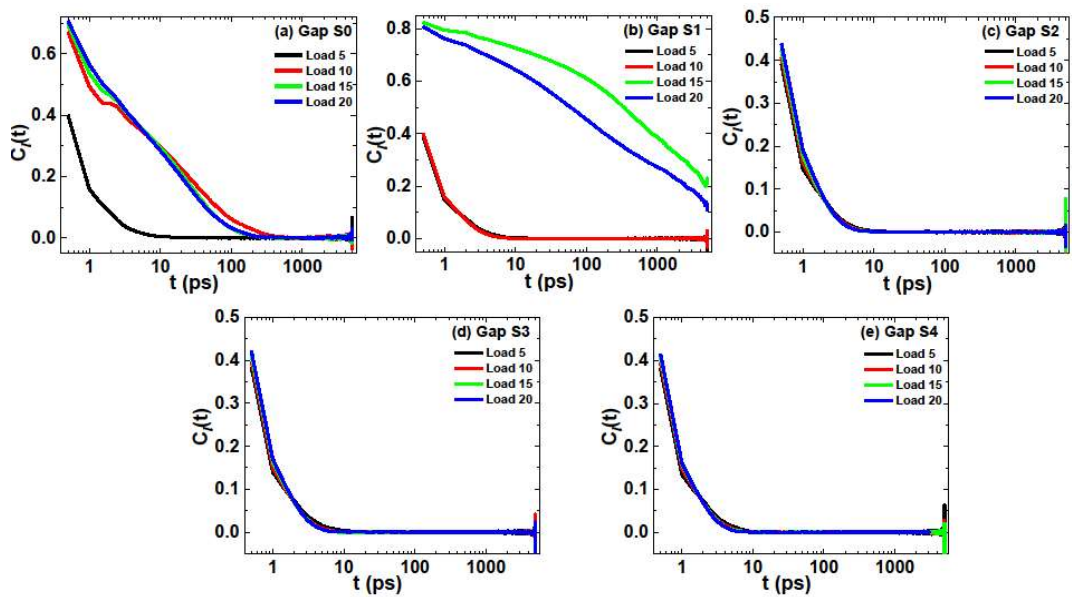

**Figure 6:** The rotational correlation function at different loadings for (a) S0, (b) S1, (c) S2, (d) S3, (e) S4.

Further, we obtained the time scale of RCF described above by fitting the RCF time dependence to an exponential decay function - $\tau$. The time scales of rotational motion obtained are shown in Figure 7. For gap S0 and S1 at lower loadings of 5 and 10 mpuc, respectively, $\tau$ shows an abrupt increase. It is likely that at these lower loadings with a smaller number of molecules, rotation is un-hindered and faster. With increase in loadings, a greater number of molecules are strongly adsorbed at the pore surface, thereby hindering the rotation motion. For higher gaps S2, S3, and S4 compared to lower gaps a smoother and almost linear increase in $\tau$ is observed, as depicted in the inset to Figure 7. This behavior is also similar to the MSD results described above.

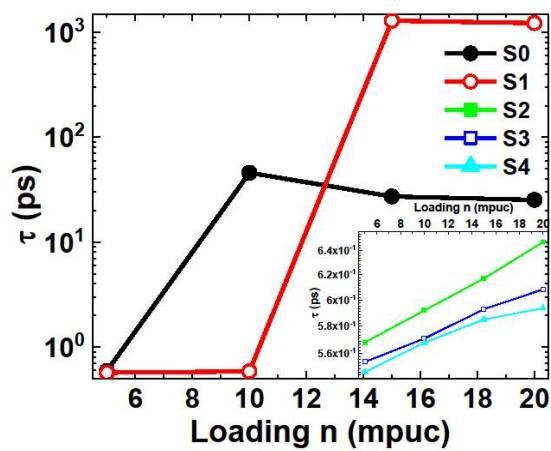

**Figure 7:** The orientational correlation times as a function of loading obtained from RCF, for inter-crystalline spacings S0 and S1, while the inset to the figure shows the same for S2, S3, and S4. Note that the inset has a linear Y-axis.

### 3.4 Effect of different exposed surfaces

Figure 8 shows the MSD comparison between different surfaces exposed for simulation cells A2, B2 and C2 along X-, Y- and Z- directions, for loadings n = 5 and 20 mpuc. For each simulation cell only one surface is exposed, such as for A2 only surface parallel to unit cell vector a is exposed, while the other two surfaces remain un-exposed. Overall, no significant difference in MSD is observed between n = 5 and 20 npuc loadings. For all the simulation cells i. e. A2, B2, and C2 and loadings, the MSD values along Z – axis exhibits the highest values as compared to X- and Y- directions. This is similar to the behavior described in Figure 4.

To further probe the $CO_2$ behavior, we separated the molecules present in crystallite pores and inter-crystalline spaces. This is particularly noteworthy for A2 and B2 surfaces, wherein the molecules within the crystallite pores and those in the inter-crystalline spaces remain isolated throughout the simulations. This is in contrast to the C2 simulation cell. This helps us to distinguish between the behavior of molecules in pores versus the inter-crystalline spacing. From this assessment, we obtain 8 molecules present in the gap and 472 in pores for A2, while for B2 16 molecules are present in the gap and 464 molecules in the pores.

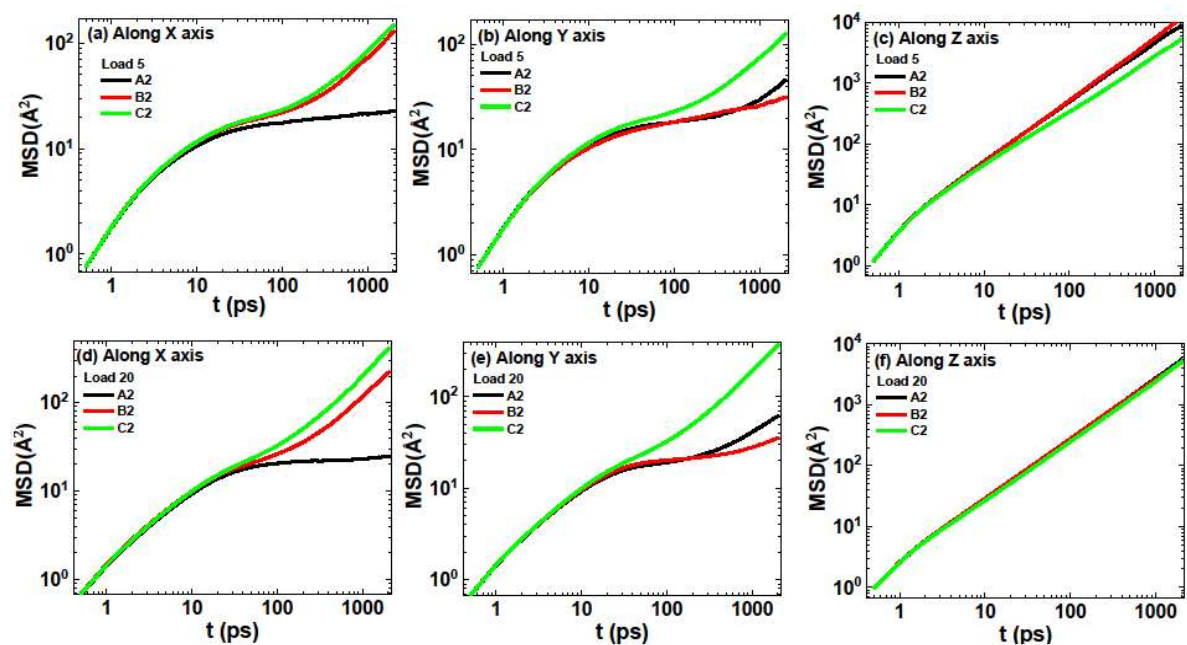

**Figure 8:** Time dependent mean square displacement (MSD) for loading n = 5 mpuc along (a) X- (b) Y- (c) Z- directions, while for n = 20 mpuc along (d) X- (e) Y- (f) Z- directions.

The corresponding Figure 9 depicts the overall MSD as a function of time for molecules in pores and inter-crystalline spaces for A2 and B2 exposed surfaces. For A2 exposed surfaces, significantly large difference between pores and gaps MSD is observed at longer times. Conversely, for B2, no significant is observed difference between the two. This behavior can be ascribed to the difference in molecular arrangement. We also calculated the ODF and OCF properties. However, no signification variation of different surfaces exposed i. e. A2, B2 and C2 is observed. This indicates that rotation remains largely unaffected by the surface.

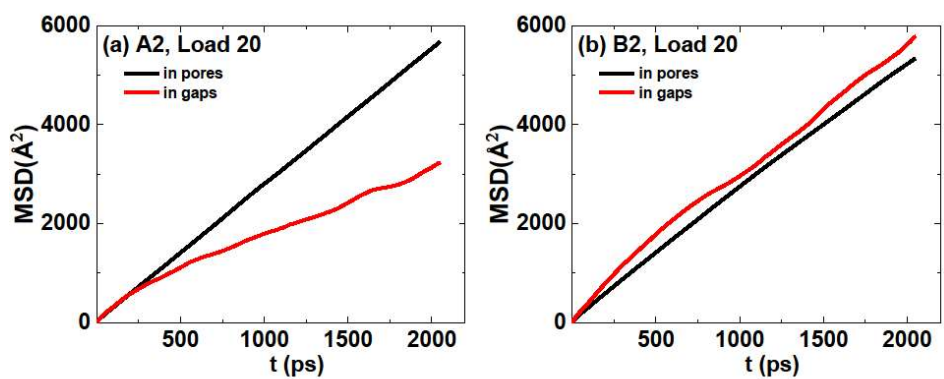

**Figure 9:** Time dependent mean square displacement (MSD) for loading n = 20 mpuc for molecules separated in pores and inter-crystalline gaps for exposed surfaces (a) A2 and (b) B2.

### 4. Discussion

Mg-MOF-74 is known to be a high-capacity $CO_2$ adsorbent. The origin of this capability lies in the existence of strong adsorption sites that are distributed along the channel-like pores in a highly ordered manner. A typical channel is seen to have 6 such strong adsorption sites. The guest-host interaction in this system is so strong that an addition of inter-crystalline spacing of 3 Å leads to the newly adsorbed molecules in the spacing staying close to the channel periphery (see Figure 2(b)). When the inter-crystalline spacing is widened, the distribution of adsorption sites in this spacing becomes delocalized. In our previous study, we concluded that the most efficient strategy to enhance the adsorption of $CO_2$ in Mg-MOF-74 is to use a sample with the smallest possible crystallites and an inter-crystalline spacing of roughly two molecular lengths (S2).[63] This is consistent with Figure 2. In the smaller inter-crystalline spacing of S1, the adsorption sites are highly localized, such that only few molecules can be adsorbed in the extra-crystallite space. However, a more homogenous adsorption site distribution present in S2-S4 can accommodate a larger amount of $CO_2$ at higher partial pressure of the guest fluid. Distribution of the adsorption sites has important consequences for the dynamics of $CO_2$ molecules. As seen in Figure 5, the diffusion coefficient of $CO_2$ in S0 and S1 decreases on an increase in loading. This is expected because of crowding at higher loading. S2-S4, however show an anomalous loading dependence of diffusion coefficient. In S0 and S1, the six adsorption sites at the pore periphery have very high intensity relative to the rest of the pore space. $CO_2$ molecules at these sites are strongly adsorbed and are localized with very little freedom to move. Addition of more $CO_2$ molecules with an increase in loading will lead to crowding and hence reduced diffusivity. In S2-S4 however, the intensity of adsorption at the six peripheral sites gets redistributed with the intensity in the pore center growing at the expense of the peripheral sites. This means that the new $CO_2$ molecules added with an increase in loadings we modeled are more likely to occupy the central regions of the pore where they experience relatively weaker interaction with the substrate and are therefore free to move. An increase in loading thus adds to the population of free and hence faster diffusing molecules, thereby increasing the overall diffusion coefficient. This explains the anomalous loading dependence of diffusivity with larger inter-crystalline spacing. This type of anomalous loading dependence of diffusion has also been observed for propane in mesoporous silica and for $SO_2$ in Na-Y zeolite.[68-70]

Although anomalous loading dependence of diffusion coefficient is observed for models with larger inter-crystalline space, the loading dependence obtained here for S0 is consistent with that obtained by Bendt et. al. and Krishna et. al.[71, 72] In absolute terms, the diffusion coefficients obtained for S0 here are comparable but slightly higher than those obtained by Bendt et. al.[71] using a DFT based force-field for $CO_2$ in a rigid Mg-MOF-74 framework. By contrast, our diffusion coefficients are slightly lower than those obtained by Krishna et. al.[72] using UFF force-field, as well as those obtained by Bendt et. al.[71] using a flexible Mg-MOF-74 framework. We note that the effects of framework flexibility on the diffusion of guest molecules in MOF are not clear. For example, while Bendt et. al.[71] found an enhancement of $CO_2$ diffusion in a flexible MG-MOF-74, framework flexibility was found to suppress the motion of benzene in MOF-5. We have therefore simulated a rigid framework.

The variation of the diffusion coefficients as a function of inter-crystalline space are different at lower and higher loadings. At a lower loading of 5 mpuc, the diffusion coefficient decreases with the introduction of inter-crystalline space, reaches a minimum in S2, and increases again to a maximum in S4. At all other loadings however, the overall diffusion coefficient increases with an increase in the inter-crystalline space. Further, comparing the overall diffusion coefficient with its Z-component, the consistent increase in the overall diffusion coefficient with inter-crystalline spacing is a result of an increase in the lateral diffusion made possible by the slab-like inter-crystalline space while the Z-component exhibits a relatively smaller change. The variation in the axial (Z-component) diffusion coefficient with inter-crystalline spacing is monotonous for all but the highest loading. This is because when inter-crystalline space is introduced the continuity of channel-like pores oriented along the Z-direction is broken and this creates bottlenecks in the axial motion leading to diffusion coefficients being affected in diverse ways. While this disruption due to a discontinuity in the channel-like pores is expected to affect mostly the Z-component of diffusion, the X and Y components should remain unaffected. Conversely, additional inter-crystalline space extending in both X and Y directions should facilitate this lateral motion in the X-Y plane. Indeed, this can be seen in Figure 5 that reveals an enhancement of the X and Y components of the diffusion coefficient with the introduction and widening of the inter-crystalline space. However, this lateral diffusion of molecules in the inter-crystalline spacing is found to be slower than the motion in the pores at least for the inter-crystalline space added above the surface A (see Figure 9). Thus, while the $CO_2$ molecules strongly adsorbed on the pore periphery in the crystallites show low mobility, the molecules in the pore center diffuse faster than those that exist in the inter-crystalline spaces. It is because of these molecules found at the pore centers that the diffusivity of $CO_3$ is enhanced with insertion of inter-crystalline space.

Loading dependence of the rotational motion is as expected – higher loadings exhibit slower rotation. The dependence of rotational time scales on the inter-crystalline space is also systematically monotonous for most loadings. As inter-crystalline space is introduced, the rotation becomes faster. This is because of the additional molecules that occupy the inter-crystalline regions where they are relatively free to rotate.

5.  Conclusions

To understand the effects of inter-crystalline space on the structural and dynamical behavior of $CO_2$ in Mg-MOF-74 we carried out an MD simulation study on $CO_2$ confined with different loadings in models of Mg-MOF-74 with systematically inserted inter-crystalline spaces of different widths. Six strong adsorption sites are found at the periphery of the pore substrate in the perfect Mg-MOF-74 substrate without any inter-crystalline space. Upon addition of inter-crystalline spacing, six additional sites of adsorption are found close to the pore opening that gradually delocalize when the inter-crystalline space is widened. This structure of adsorbed $CO_2$ is found to significantly affect the dynamical behavior of the guest molecules. The translational diffusivity of the guest molecules in the substrate model without inter-crystalline space (S0) decreases as the loading is increased, as reported for other studies, whereas the loading dependence of diffusivity in Mg-MOF-74 models with wider inter-crystalline space is anomalous. This behavior is a result of the growth of guest population at the pore center at the expense of the strongly adsorbed and hence less mobile population found at the pore periphery. In general, addition of inter-crystalline space is found to enhance both translational as well as rotational motion of guest molecules.

**Conflict of interest statement**

The authors declare that there is no conflict of interest.

**Acknowledgements**

I. D. acknowledges KIFÜ for awarding the access to resource based in Hungary at Budapest. S. G. and D. R. C. are supported by the U.S. Department of Basic Energy, Office of Science, Office of Basic Energy Sciences, Division of Chemical Sciences, Geosciences and Biosciences, Geosciences Program, grant number DESC0006878.

**Contributions**

Conceptualization, methodology, validation, formal analysis, investigation, writing original draft, project administration, I. D. and S. G.; Resources, data curation, visualization, I. D. Software and coding, S. G.; Review and editing, I. D., M. C. B., S. G., and D. R. C.